\documentclass[%
superscriptaddress,
nolongbibliography,
 amsmath,amssymb,
 aps,
pre,
]{revtex4-2}
\usepackage{graphicx}
\usepackage{xcolor}

\begin{document}


\title{Wake-Tail Effects in Two-Dimensional Wave Refocusing}


\author{Theodoros T. Koutserimpas}
\email[]{tkoutserimpas@mail.ntua.gr}
\affiliation{School of Electrical and Computer Engineering, National Technical University of Athens, Athens GR-15780, Greece}


\date{\today}

\begin{abstract}
In even spatial dimensions, solutions of the wave equation violate Huygens’ principle, producing a persistent wake tail inside the light cone rather than a sharply localized propagating front. This intrinsic tail complicates refocusing. Here, we examine how the wake-tail structure of the two-dimensional wave equation affects refocusing, using the analytically tractable example of a pulse generated by a source localized in both space and time. Two idealized concentration strategies are considered. A spatial mirror reflects the outgoing pulse and produces refocusing, but the redirected signal is broadened, with the wake tail preserving its causal ordering behind the propagating front. A second strategy employs a time mirror generated by abrupt temporal modulation of the phase velocity, producing temporal reflection and transmission. This mechanism introduces an anti-causal response of the wake-tail, reversing its temporal ordering in a time-reversal-like manner; however, the pulse still undergoes distortion and wake-tail contributions persist through secondary radiation at the refocus point. These results demonstrate the fundamental connection between Huygens’ principle and wave concentration, showing that the wake-tail structure intrinsic to two-dimensional propagation imposes a fundamental limit on perfect refocusing, even under idealized conditions.
\end{abstract}


\maketitle

\section{Introduction\label{intro}}

Refocusing plays a central role in many areas of wave physics and engineering, enabling imaging, and energy localization by usually exploiting the time-reversal symmetry of the wave equation in lossless media. Such techniques have found applications across acoustics, electromagnetics, medical imaging, nondestructive evaluation, and wireless communications, where recorded wavefields are time reversed and re-emitted to refocus energy back at the original source location \cite{Fink1992_I,Fink1992_II,Fink1992_III,Fink1997,Blomgren2002,Lerosey2004}.

However, the effectiveness of refocusing depends not only on practical considerations such as losses, finite apertures, or incomplete measurements, but also on more fundamental properties of wave propagation. A particularly important and less frequently discussed limitation arises from the dimensional structure of the wave equation itself. In odd spatial dimensions, such as three-dimensional free space, solutions satisfy Huygens’ principle: disturbances propagate on sharply defined wavefronts, and once the front passes, the field vanishes. In contrast, in even spatial dimensions the wave equation violates Huygens’ principle, and impulsive sources generate persistent fields inside the light cone, referred to as wake tails. As a result, energy continues to arrive at observation points after the main front has passed, producing temporal memory in the propagating field \cite{Hadamard,CourantHilbert1962,Garabedian1964}. 

The intrinsic wake-tail structure of two-dimensional wave propagation imposes fundamental limitations on wave refocusing. In contrast to odd-dimensional systems, where energy remains sharply localized at the propagating front, two-dimensional waves exhibit persistent trailing contributions inside the light cone. As a result, the refocused signal cannot remain perfectly concentrated, but instead undergoes temporal and spatial broadening due to the continuous wake-tail contributions. Consequently, even idealized refocusing mechanisms are required to reconstruct not only the propagating front, but also the extended interior field distribution generated during propagation, making exact refocusing fundamentally unattainable in the presence of wake-tail effects.

In this work, we investigate how the wake-tail structure inherent to two-dimensional wave propagation affects wave refocusing. To isolate the fundamental mechanisms involved, we consider a minimal and analytically tractable scenario in which a pulse is generated by an impulsive source localized in both space and time, and we analyze how different refocusing strategies reconstruct the field. Two ideal representative mechanisms are examined: spatial reflection from a perfect mirror and temporal reflection induced by an abrupt modulation of the phase velocity. While both mechanisms produce wavefront refocusing, neither reproduces the exact pulse at the refocus point, revealing intrinsic limitations associated with two-dimensional propagation.

The results highlight how geometric properties of wave propagation alone, independent of losses or experimental imperfections, can limit perfect refocusing. Understanding these effects is important for interpreting experiments and designing wave-control strategies in effectively two-dimensional systems.

\section{The wake-tail effect}

In this section, we briefly review the role of dimensionality in the wave equation. To do so, we consider the retarded Green function, which satisfies

\begin{equation}
\left( {\frac{1}{{{c^2}}}\partial _t^2 - {\Delta _d}} \right){g_d}(r,t) = \delta (t){\delta ^{(d)}}({\bf{r}}),
\label{wave-eq}
\end{equation}
where $\Delta_d$ is the Laplacian in $d$ spatial dimensions, $c$ denotes the phase velocity, $r=|{\bf{r}}|$ and $\delta$ is the Dirac distribution. Causality is assumed, i.e., $g_{d}=0$ for $t<0$.

For the $d=1$ case, the solution is straightforward \cite{Barton,Kythe,Rother}: 

\begin{equation}
{g_1}(r,t) = \frac{1}{2}H\left( {t - \frac{{\left| x \right|}}{c}} \right),
\label{Green1}
\end{equation}
where $H(x)$ is the Heaviside distribution. The one-dimensional Green function give rise to a particularly unusual behavior. Although the source is an impulsive excitation that exists only for an instant in time, it generates a wave whose effect persists throughout the region behind the propagating front. In one dimension there is no geometric spreading and therefore no free-space amplitude decay; consequently, the field remains constant after the front has passed. As a result, once the wavefront propagates through the medium, the entire accessible region behind it remains filled with the disturbance.

For $d=2$ the solution is \cite{Barton,Kythe,Rother}:

\begin{equation}
{g_2}(r,t) = \frac{{H(t - {r \mathord{\left/
 {\vphantom {r c}} \right.
 \kern-\nulldelimiterspace} c})}}{{2\pi \sqrt {{t^2} - {{\left( {{r \mathord{\left/
 {\vphantom {r c}} \right.
 \kern-\nulldelimiterspace} c}} \right)}^2}} }}.
\label{Green2}
\end{equation}

An interesting feature of the two-dimensional Green function emerges when compared with the corresponding solution in one dimension. In two dimensions, the propagator decays as $(t-\tau)^{-1}$, so that beyond the abrupt arrival at $t-\tau = r/c$, the solution produces a wake tail that, in principle, persists indefinitely. As discussed later, this behavior arises from the geometry of wave propagation in different spatial dimensions. 

For $d=3$ the solution is \cite{Barton,Kythe,Rother}:

\begin{equation}
{g_3}(r,t) = \frac{{\delta (t - {r \mathord{\left/
 {\vphantom {r c}} \right.
 \kern-\nulldelimiterspace} c})}}{{4\pi r}}.
\label{Green3}
\end{equation}

In contrast, the three-dimensional propagator describes a propagating front whose amplitude decays proportionally to $r^{-1}$ due to geometric spreading, corresponding to free-space losses. However, the three-dimensional geometry reveals no temporal memory: no wake tail is produced, and the duration of the observed pulse matches that of the source. 

Although most physical systems involve wave propagation in one, two, or three spatial dimensions, the Green functions of the wave equation can be systematically constructed for arbitrary dimension. In particular, higher-dimensional solutions may be found from the recursive relation

\begin{equation}
{g_{d + 2}}(r,t) =  - \frac{1}{{2\pi r}}\frac{\partial }{{\partial r}}{g_d}(r,t),
\label{recursive}
\end{equation}
following relevant analysis in \cite{Friedlander,Rother} (see also Appendix (A) for a proof). Starting from the known solutions in $d=2$ and $d=3$, this relation reveals a fundamental qualitative difference between even and odd spatial dimensions. The recursion is not applied to $d=1$, since it assumes radial symmetry; the one-dimensional Green function is too singular at $x=0$ for the operator to be applied directly. 

In even dimensions, repeated application of the recursion produces terms of the general form $\sim H(t-r/c)[t^2-(r/c)^2]^{-(d-1)/2}$, so that the solution retains support throughout the interior of the light cone. The exponent remains half-integer and does not collapse to a delta distribution under differentiation, resulting in persistent wake tails. This behavior directly implies a violation of Huygens' principle in even spatial dimensions, since the field does not propagate solely on the wavefront but instead leaves a nonvanishing contribution throughout the interior region \cite{Hadamard,CourantHilbert1962,Garabedian1964}. In contrast, in odd dimensions and for $d \ge 3$, successive radial derivatives eventually convert the solution into delta distributions and their derivatives supported strictly on the light cone. Consequently, the propagator is supported only on the wavefront itself, and Huygens' principle holds.

From a spectral viewpoint, odd dimensions are associated with pole-type singularities that produce sharp propagating fronts, whereas even dimensions lead to branch-cut behavior in frequency space, resulting in temporal memory, i.e., wake tails. In physical terms, wave propagation redistributes energy differently depending on spatial dimension. In three dimensions, energy spreads over a spherical surface of area $4\pi r^2$, while in two dimensions it spreads over a circular front of circumference $2\pi r$. Because geometric spreading is weaker in even dimensions, energy cannot remain confined to a propagating shell and instead continuously fills the interior region behind the front. This geometric mechanism ultimately explains both the violation of Huygens' principle and the emergence of temporal tails, as well as the associated challenges for refocusing in even-dimensional wave propagation.

Next, we examine the limitations inherent to two-dimensional wave propagation and investigate how refocusing effects can be generated using two idealized configurations: a spatial mirror and a temporal mirror. In particular, we analyze how these mechanisms influence the refocusing process in the presence of the intrinsic wake tail characteristic of two-dimensional propagation.

\section{\label{spatial_mirror} Wave Refocusing in 2D: Spatial Mirror Approach
}

We consider the wave generated by an impulsive source located at the origin. To refocus the wave, we introduce a perfectly reflecting circular mirror at radius $r=a$, imposing a Dirichlet boundary that enforces complete reflection. The resulting initial--boundary-value problem reads
\begin{equation}
\frac{1}{{{c^2}}}\frac{{{\partial ^2}{u_{sp}}}}{{\partial {t^2}}} 
-  \Delta u_{sp} 
= \frac{{\delta (r)\delta (t)}}{{2\pi r}},
\label{spatial-mirror}
\end{equation}
together with the boundary condition ${u_{sp}}(a,t) = 0$ and causal initial conditions ${u_{sp}}(r,0^-) = {\partial_t}{u_{sp}}(r,0^-) = 0$, where $\Delta = \Delta_2$. Taking the Laplace transform of Eq.~(\ref{spatial-mirror}) with respect to time yields
\begin{equation}
\frac{{{s^2}}}{{{c^2}}}{U_{sp}} 
- \frac{{{\partial ^2}{U_{sp}}}}{{\partial {r^2}}} 
- \frac{1}{r}\frac{{\partial {U_{sp}}}}{{\partial r}} 
= \frac{{\delta (r)}}{{2\pi r}},
\label{Laplace-spatial}
\end{equation}
where $U_{sp}(r,s)$ denotes the Laplace transform of ${u_{sp}}(r,t)$. Enforcing the boundary at $r=a$, we expand
\[
{U_{sp}}(r,s) = \sum_{n = 1}^\infty A_n\, J_0\!\left(\frac{k_n r}{a}\right),
\]
where $J_0$ is the Bessel function of the first kind of order zero and $k_n$ denotes the $n$th positive root of $J_0(k)=0$.

The radial delta distribution admits the Bessel--Fourier expansion \cite{Watson}
\begin{equation}
\frac{{\delta (r)}}{{2\pi r}} 
= \frac{1}{{\pi a^2}}
\sum_{n = 1}^\infty 
\frac{J_0\!\left(\frac{k_n r}{a}\right)}{J_1^2(k_n)},
\label{delta-Bessel}
\end{equation}
where $J_1$ is the Bessel function of the first kind of order one. Matching coefficients yields
\begin{equation}
{U_{sp}}(r,s) 
= \frac{{c^2}}{{\pi a^2}}
\sum_{n = 1}^\infty 
\frac{J_0\!\left(\frac{k_n r}{a}\right)}
{\left(s^2 + c^2\frac{k_n^2}{a^2}\right) J_1^2(k_n)}.
\label{solution-Laplace}
\end{equation}
Inverting term-by-term using standard Laplace tables gives
\begin{equation}
{u_{sp}}(r,t) 
= \frac{c^{2}t\,H(t)}{\pi a^2}
\sum_{n = 1}^\infty 
\frac{J_0\!\left(\frac{k_n r}{a}\right)}{J_1^2(k_n)}
\,\mathrm{sinc}\!\left(\frac{c k_n t}{a}\right),
\label{solution}
\end{equation}
which is a complete modal representation of the field in the disk (such representation is commonly referred to as a modal decomposition; see, e.g., \cite{Morse}). While Eq.~(\ref{solution}) is mathematically explicit, its physical content (in particular, how refocusing emerges and when it occurs) is not immediately transparent. To make the refocusing mechanism explicit, it is useful to reorder the Laplace-domain series in Eq.~(\ref{solution-Laplace}) into a free-space contribution plus a boundary-induced correction.

Using standard summation identities for Bessel eigenexpansions in a disk, Eq.~(\ref{solution-Laplace}) can be written equivalently as
\begin{equation}
{U_{sp}}(r,s)
= \frac{{c^2}}{{\pi a^2}}\sum\limits_{n = 1}^\infty
\frac{{J_0\!\left(\frac{k_n r}{a}\right)}}{\left(s^2 + c^2\frac{k_n^2}{a^2}\right)J_1^2(k_n)}
= \frac{1}{2\pi}K_0\!\left(\frac{rs}{c}\right)
-\frac{1}{2\pi}\frac{K_0\!\left(\frac{as}{c}\right)}{I_0\!\left(\frac{as}{c}\right)}\,I_0\!\left(\frac{rs}{c}\right),
\label{alt-solution-Laplace}
\end{equation}
where $I_0$ and $K_0$ are the modified Bessel functions. This decomposition has a direct physical interpretation. The first term is precisely the Laplace-domain free-space propagator in two dimensions; it represents the outward-propagating field launched by the impulsive source. The second term is a homogeneous solution added to enforce the boundary condition at $r=a$, and it encodes the field generated by reflection at the mirror. Importantly, this reflected contribution ``turns on'' only after the minimum round-trip travel time, thereby revealing the refocusing time scale.

Taking the inverse Laplace transform of Eq.~(\ref{alt-solution-Laplace}) yields
\begin{equation}
{u_{sp}}(r,t)
=
\frac{{H(t - {r \mathord{\left/
 {\vphantom {r c}} \right.
 \kern-\nulldelimiterspace} c})}}{{2\pi \sqrt {{t^2} - {{\left( {{r \mathord{\left/
 {\vphantom {r c}} \right.
 \kern-\nulldelimiterspace} c}} \right)}^2}} }}
+\frac{{c\,H(ct + r - 2a)}}{{a\pi }}
\sum\limits_{n = 1}^\infty
\frac{J_0\!\left(\frac{k_n r}{a}\right)}{k_n J_1^2(k_n)}
\sin\!\left(\frac{c k_n t}{a}\right).
\label{alt-solution}
\end{equation}

The first term is the usual two-dimensional free-space response \eqref{Green2}. The second term is the mirror-induced contribution and contains the explicit causal factor $H(ct+r-2a)$: for a point at radius $r$, the boundary cannot influence the field until a signal can reach the mirror and return, implying the earliest reflected arrival at $t=(2a-r)/c$. In particular, at the source point $r=0$ the reflected field turns on at the refocusing time $t_f=\frac{2a}{c}.$ Equation~(\ref{alt-solution}) therefore makes the refocusing mechanism transparent: the spatial mirror generates a delayed, inward-propagating contribution that concentrates energy back toward the origin at $t_f$. However, because two-dimensional propagation intrinsically produces wake tails, the reflected field retains causal temporal ordering and its wake-tail is preceded by its wave front. Therefore, refocusing is temporally extended and imperfect. 

\begin{figure}
\includegraphics{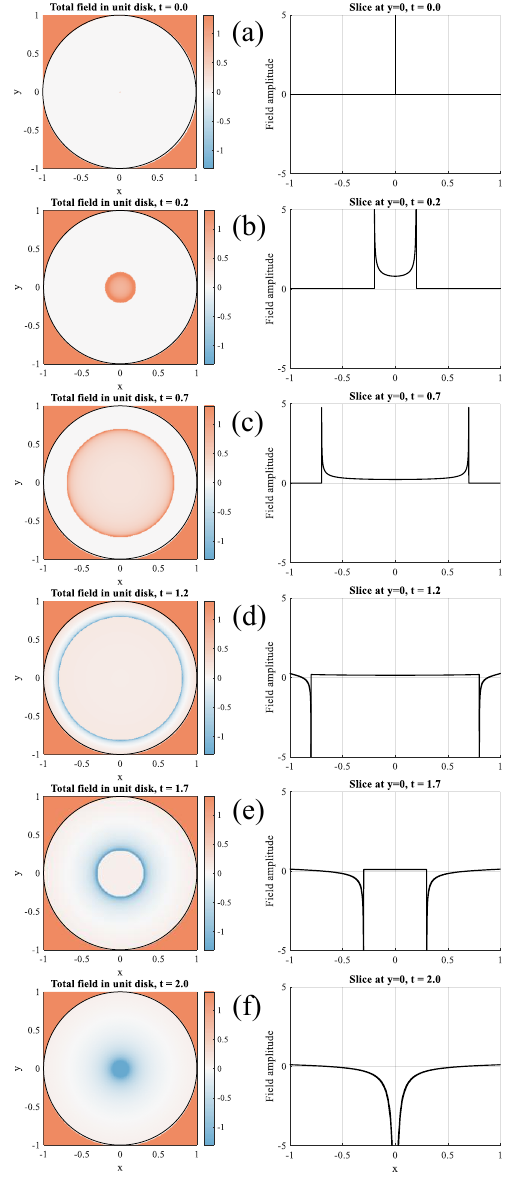}
\caption{
Chronophotographs of wave excitation at the center and the subsequent refocusing within a perfectly reflecting circular boundary of unit radius ($a=1$) with phase velocity $c=1$. Each snapshot displays the total field together with the corresponding slice along $y=0$ at (a) $t=0$, (b) $t=0.2$, (c) $t=0.7$, (d) $t=1.2$, (e) $t=1.7$, and (f) $t=2$. For $t<1$, only the initial pulse is present, propagating outward while leaving behind the characteristic two-dimensional wake tail. After reflection at the boundary (occurring for $t>1$), an inward-propagating contribution appears and progressively redirects energy toward the origin. Refocusing becomes apparent at $t=2a/c=2$; however, the reconstructed field remains temporally extended due to the intrinsic wake-tail structure of two-dimensional wave propagation, preventing perfect wave concentration. Consequently, the field at the refocusing time $t=2$ (panel f) is never as spatially concentrated as the initial excitation at $t=0$ (panel a). The movie associated with the chonophotographs of Fig. \ref{Fig:Temporal_Mirror} can be found as Supplementary Material \cite{Sup}.
}
\label{Fig:Spatial_Mirror}
\end{figure}

Figure~\ref{Fig:Spatial_Mirror} shows chronophotographs of the wave field generated by an impulsive excitation at the origin inside a circular domain with a perfectly reflecting boundary of unit radius $a=1$, for a medium with phase velocity $c=1$ and various times. For each representative time, the two-dimensional panel displays the total field inside the disk and the corresponding slice plot along $y=0$.

At early times, Fig.~\ref{Fig:Spatial_Mirror}(b),(c) where $t<a/c=1$ ($t=0.2$ and $t=0.7$), the field consists solely of the direct wave produced by the impulsive source. The disturbance propagates radially outward while leaving behind a nonzero field throughout the interior region already traversed by the front, in perfect agreement with the theoretical analysis. This behavior reflects the violation of Huygens' principle in two dimensions, where the propagator generates a persistent wake tail rather than a sharply localized propagating shell. The slice plots confirm that the field remains finite behind the front instead of collapsing to a thin wavefront.

Once the outward wave reaches the boundary at $r=1$ and $t=1$, reflection occurs and an inward-propagating component is generated. This contribution becomes visible in the later chronophotographs at Fig.~\ref{Fig:Spatial_Mirror}(d),(e) ($t=1.2$ and $t=1.7$), where the two-dimensional plots show energy returning toward the center. This behavior is consistent with the analytical solution, where the mirror-induced contribution is activated by the causal factor $H(ct+r-2a)$, representing the earliest return of reflected signals. Importantly, the returning pulse also retains a wake tail which is preceded by its sharp wavefront.

At the round-trip time $t=2a/c=2$, Fig.~\ref{Fig:Spatial_Mirror}(f), the returning energy concentrates near the origin, albeit with noticeable spreading, resulting in only partial refocusing of the initial disturbance shown in Fig.~\ref{Fig:Spatial_Mirror}(a). The chronophotographs further indicate that the reconstructed signal remains temporally extended. The inward-propagating contribution preserves the characteristic two-dimensional wake-tail structure, thereby preventing the formation of a sharply localized refocusing pulse.

Figs. (\ref{Fig:Spatial_Mirror}) (a), (f) illustrate the central theoretical result: although a spatial mirror redirects energy back toward the source and produces refocusing at the expected round-trip time, the intrinsic wake-tail behavior of two-dimensional wave propagation prevents perfect concentration at the center. Refocusing occurs, but the reconstructed pulse is broadened and cannot reproduce the original impulsive excitation exactly.

In the mirror-induced refocusing considered here, the wake tail follows the returning wavefront, whereas perfect time-reversal reconstruction would require a fundamentally anti-causal ordering, in which the tail precedes the front. This mismatch in temporal structure further inhibits ideal refocusing, even under perfectly reflecting conditions. In the next section, we demonstrate that a temporal mirror, realized through an abrupt temporal modulation of the propagation velocity, can partially emulate this reversed ordering by generating a returning wave in which the tail precedes the front, thereby more closely approximating the conditions necessary for time-reversal reconstruction.

\section{\label{spatial_mirror} Wave Refocusing in 2D: Temporal Mirror Approach
}

In this approach, refocusing is achieved through an abrupt temporal change of the material properties \cite{Bacot,Bal}. The initial excitation remains the same, namely the impulse response produced by a point source at the origin at time $t=0$. The difference here is that, at time $\tau_0=a/c>0$, the phase velocity undergoes an infinitesimal-duration perturbation ($\Delta t \to 0$) with modulation depth $\alpha$, $c(t)=c_0 \sqrt{(1+\alpha \delta(t-\tau_0))}$. This situation is modeled by the wave equation

\begin{equation}
\frac{1}{{c_0^2(1 +\alpha \delta (t - {\tau _0}))}}\frac{{{\partial ^2}{u_{tm}}}}{{\partial {t^2}}} - \Delta u_{tm}  = \frac{{\delta (r)\delta (t)}}{{2\pi r}}.
\label{time-mirror}
\end{equation}

with $\mathop {\lim }\limits_{r \to \infty } {u_{tm}}(r,t) = 0$ and causal initial conditions ${u_{tm}}(r,0^-) = {\partial_t}{u_{tm}}(r,0^-) = 0$.

Due to this sudden temporal change of the material properties, the wave undergoes temporal reflection and transmission \cite{Morgenthaler}. For $0<t<\tau_0$, the solution coincides with that obtained from the initial excitation pulse in the previous approach (\ref{Green2}). The solution is modified only after the velocity modulation occurring at $t=\tau_0$.  

We next determine the field conditions at $t=\tau_0$. Since the governing equation contains no singular source acting directly on $u_{tm}$ at that time, the field itself remains continuous,

\begin{equation}
u_{tm}(\tau_0^+) = u_{tm}(\tau_0^-).
\label{boundary-relations1}
\end{equation}

To obtain the jump condition for the time derivative, we first linearize the modulation term, ${\left( {1 + \alpha \delta (t - {\tau _0})} \right)^{ - 1}} \approx 1 - \alpha \delta (t - {\tau _0})$ (first-order Born approximation) and note that, in a neighborhood of $\tau_0$, the right-hand side of \eqref{time-mirror} vanishes since no point source is present at that time. The equation therefore becomes

\begin{equation}
\frac{1}{{{c_0^2}}}\partial _t^2{u_{tm}} - \Delta {u_{tm}} = \frac{\alpha }{{{c_0^2}}}\delta (t - {\tau _0})\partial _t^2{u_{tm}}(\tau _0^ - ).
\label{boundary}
\end{equation}

Integrating \eqref{boundary} over the interval
$[\tau_0-\varepsilon/2,\tau_0+\varepsilon/2]$ and assuming $\Delta u_{tm}$ remains bounded yields

\begin{equation}
{\partial _t}{u_{tm}}(\tau _0^ + ) - {\partial _t}{u_{tm}}(\tau _0^ - ) = \alpha \partial _t^2{u_{tm}}(\tau _0^ - ).
\label{boundary-relations2}
\end{equation}

Equations \eqref{boundary-relations1} and \eqref{boundary-relations2} constitute the continuity relations associated with temporal transmission and reflection. They effectively define explicit time-domain operators acting on the field across the temporal interface. The corresponding time-domain transmission and reflection operators are $\mathcal{T}f(t) = H(t - \tau_0)\left( f(t) + \tfrac{\alpha}{2}\,\partial_t f(t) \right)$
and $\mathcal{R}f(t) = H(t - \tau_0)\, \tfrac{\alpha}{2}\,\partial_t f(2\tau_0 - t)$, whose derivation is provided in Appendix~B.

A special treatment is, however, required for the reflected pulse. Due to the presence of the wake tail in two dimensions, portions of the reflected field for $t>\tau_0$ propagate back toward the origin. After reaching the refocus point, these components act as an effective secondary excitation and subsequently re-propagate outward. Consequently, an additional contribution must be included, corresponding to the outward field generated by this refocusing process, namely $u_{ref}(0,t)$. Determining this term constitutes an inverse problem: one seeks the effective source $s(t)$ that produces a field $h(t)=u_{ref}(0,t)$ propagating outward in two-dimensional free space. Using Green-function analysis leads to a singular Abel-type convolution:

\begin{equation}
h(t) = \int_0^t {\frac{{s(\tau )}}{{2\pi (t - \tau )}}} .
\label{source-relation}
\end{equation}

The resulting field is:

\begin{equation}
u_{tm}(r,t)
=
g(r,t)
+
H(t-\tau_0)\left[
\frac{\alpha}{2}\,\partial_t g(r,t)
+
\frac{\alpha}{2}\,\partial_t g\!\left(r,2\tau_0-t\right)
\right]
+
\int_{0}^{t} s(\tau)\, g(r,t-\tau)\, d\tau .
\label{solution-time-mirror}
\end{equation}

where $g$ denotes the impulse response generated at the origin at time $t=0$, given by \eqref{Green2} and $h(t) = \frac{\alpha}{4\pi}
\left[
\frac{H(t - \tau_0)\, H(2\tau_0 - t)}{(2\tau_0 - t)^2}
\right]$, which is used to numerically calculate $s$ from \eqref{source-relation}.

\begin{equation}
\begin{aligned}
u_{tm}(r,t)
&= \frac{H\!\left(t - \frac{r}{c}\right)}
        {2\pi \sqrt{t^2 - \left(\frac{r}{c}\right)^2}} \\
&\quad
- H(t-\tau_0)\frac{\alpha}{2}
\Bigg[
\frac{t\, H\!\left(t-\frac{r}{c}\right)}
     {2\pi \left(t^2-\left(\frac{r}{c}\right)^2\right)^{3/2}}
\\
&\qquad\qquad
-
\frac{(2\tau_0 - t)\,
      H\!\left(2\tau_0 - t - \frac{r}{c}\right)}
     {2\pi
      \left((2\tau_0 - t)^2 - \left(\frac{r}{c}\right)^2\right)^{3/2}}
\Bigg]
\\
&\quad
+ \int_0^t
s(\tau)\,
\frac{H\!\left(t-\tau-\frac{r}{c}\right)}
     {2\pi \sqrt{(t-\tau)^2-\left(\frac{r}{c}\right)^2}}
\, d\tau .
\end{aligned}
\label{solution-time-mirror-regular}
\end{equation}

Since $h(t) \propto (2{\tau _0} - t)^{-2}$, the associated inverse problem is particularly challenging, as it becomes unstable in the vicinity of $t = 2\tau_0$ (the refocus time). Numerically, the evaluation of the last term essentially reduces to determining the source $s(t)$. To obtain a stable solution, we employ a finite-radius matching at $r=b$, imposing $u_{ref}(b,t)=h(t-b/c)$, as described in Appendix~C.

The time derivatives of the propagator contain contributions supported on the light cone that originate from differentiating the Heaviside factor $H(t-r/c)$. These terms have been omitted. 

In any realizable setting, the excitation and the temporal modulation have finite duration and finite bandwidth, and the measured field is effectively convolved with the instrument response. This regularization smooths the cone singularity and replaces the $\delta$-supported wavefront by a finite-amplitude, finite-width transient. Since our interest here is the refocusing dynamics of the wake-tail field inside the light cone (which remains integrable and physically observable), we report the regular part of the  solution and omit the explicit cone-supported singular contributions for clarity. This omission does not affect the causal support, the refocusing time, or the leading-order amplitude of the partially time-reversed component.

\begin{figure}
\includegraphics{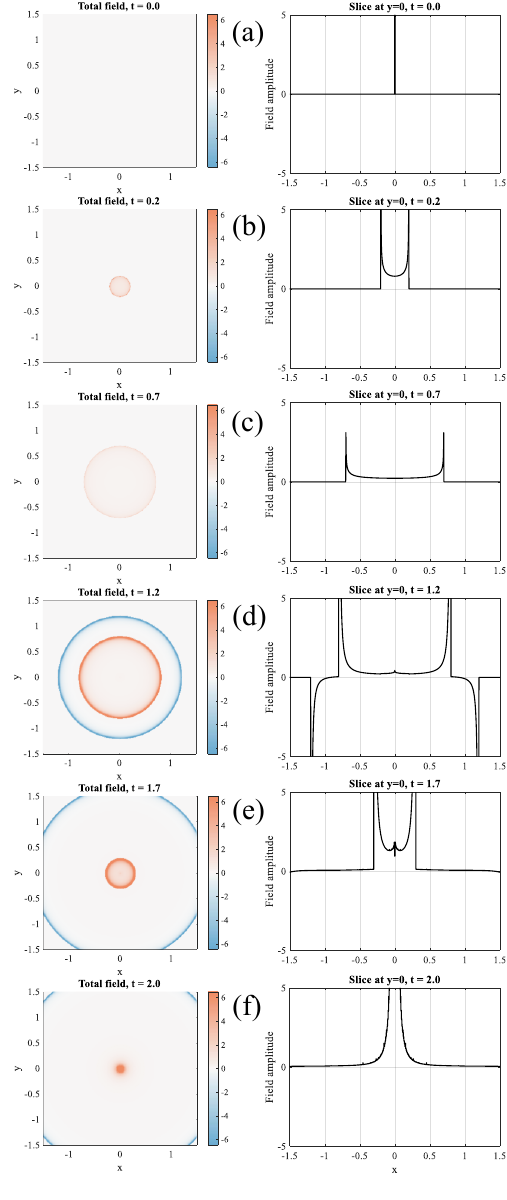}
\caption{
Chronophotographs of the wave field generated by an impulsive excitation at the center. At $\tau_0 = 1$, a temporal modulation induces temporal reflection and transmission. The medium parameters are $c_0=1$ and modulation depth $\alpha = 1$. Each snapshot shows the total field together with the corresponding slice along $y=0$ at (a) $t=0$, (b) $t=0.2$, (c) $t=0.7$, (d) $t=1.2$, (e) $t=1.7$, and (f) $t=2$. For $t < 1$, only the initial pulse is present, propagating outward while leaving behind the characteristic two-dimensional wake tail. After the temporal modulation at $t=1$, temporal reflection and transmission generate an inward-propagating contribution that progressively redirects energy toward the origin and an outward-propagating component continues to diverge from the center, respectively. Refocusing becomes apparent at the round-trip time $t = 2\tau_0 = 2$. As in Fig.~\ref{Fig:Spatial_Mirror}, the reconstructed field remains temporally extended due to the intrinsic wake-tail structure of two-dimensional wave propagation, preventing perfect wave concentration. In contrast to the spatial mirror case, the refocusing pulse now displays anti-causal temporal ordering, resembling time-reversal, with the wake tail preceding the wavefront, as enforced by temporal reflection. The movie associated with the chonophotographs of Fig. \ref{Fig:Temporal_Mirror} can be found as Supplementary Material \cite{Sup}.
}
\label{Fig:Temporal_Mirror}
\end{figure}

Figure~\ref{Fig:Temporal_Mirror} illustrates the temporal evolution of the wave field generated by an impulsive excitation at the origin in a homogeneous two-dimensional medium with $c_0=1$, together with the corresponding field slices along \(y=0\). A temporal modulation at \(t=\tau_0=1\) produces temporal reflection and transmission, enabling refocusing of the wave energy at later times.

At early times Figs.~\ref{Fig:Temporal_Mirror} (b) and (c) associated with $t=0.2$ and $t=0.7$, only the directly excited field is present. The disturbance propagates radially outward while leaving a nonvanishing field inside the region already traversed by the wavefront, forming the characteristic wake tail associated with the violation of Huygens' principle in two dimensions. The slice plots clearly show that the field is not localized at a sharp front but remains distributed throughout the interior region.

After the temporal modulation at \(t=1\), Figs.~\ref{Fig:Temporal_Mirror} (d) and (e) reveal the appearance of a temporally reflected component propagating inward, while the transmitted component continues to propagate outward. The inward component progressively redirects energy toward the origin, producing an increasing concentration of energy near the center, while the outgoing wave continues to expand toward the boundaries. Importantly, the reflected contribution displays  anti-causal temporal ordering: the wake tail now precedes the sharp front. This behavior demonstrates that a temporal mirror resembles time-reversal dynamics. Nevertheless, the pulse is not an exact time-reversed replica of the original excitation, since the reflection operator inherently involves a time-derivative action on the field. \cite{Bal} develops a rigorous mathematical framework for wave equations with time-singular coefficients, establishing regularity and energy estimates together with quantitative refocusing results that characterize the extent to which the reconstructed wave approximates ideal time reversal. In the present work, however, our primary focus lies on the wake-tail phenomena showcased by two-dimensional wave solutions and on how these intrinsically limit the refocusing process.

At the round-trip time \(t = 2\tau_0 = 2\), Fig.~\ref{Fig:Temporal_Mirror} (f), energy refocuses at the origin. However, the reconstructed signal remains temporally broadened due to the intrinsic wake-tail structure of two-dimensional propagation, preventing perfect recovery of the original impulsive excitation. Thus, although temporal modulation enforces the anti-causal time ordering of the reconstructed signal, perfect refocusing remains unattainable because of the fundamental propagation properties of two-dimensional waves. Additionally, the wake-tail contribution implies that once the temporally reflected field reaches the origin it does not simply ``stop'': the nonzero trace at the refocus point, \(r=0\), acts as an effective secondary excitation that re-radiates outward. In our formulation this secondary radiation is captured by the convolution term in \eqref{solution-time-mirror}, where the source function \(s(t)\) is chosen so that the radiated field reproduces the prescribed refocusing trace \(h(t)\). Recovering \(s(t)\) from \eqref{source-relation} is an Abel-type inverse problem of the first kind and is inherently ill-conditioned because the kernel is weakly singular and smoothing. As a result, small numerical errors (or discretization noise) are strongly amplified, especially near the refocusing time \(t\approx 2\tau_0\) where the target trace varies rapidly. The finite-radius matching and Tikhonov regularization used in the simulations should therefore be viewed as physically motivated regularizations rather than as purely technical fixes.

\section{Discussion}
\label{sec:discussion}

In this work we examined a fundamental limitation of refocusing that arises solely from the dimensional structure of the wave equation. In two spatial dimensions the retarded Green function has support throughout the interior of the light cone, producing a persistent wake tail that carries temporal memory. As a consequence, any refocusing mechanism must reconstruct not only a propagating wavefront, but also the correct temporal content and ordering of the tail. This requirement is intrinsically more demanding than in odd-dimensional settings (e.g., three-dimensional free space), where Huygens' principle holds and the propagator is supported only on the wavefront. Using an impulsive point excitation as a minimal and analytically tractable test case, we compared two representative refocusing strategies. 

The spatial mirror approach (Dirichlet reflection at a circular boundary) produces refocusing at the expected round-trip time, but the reconstructed field remains temporally extended and never collapses to a sharply localized replica of the original impulse response. The chronophotographs also reveal that the reflected pulse retains causal temporal ordering: the returning wavefront is followed by its wake tail.

The temporal mirror approach, implemented through an instantaneous modulation of the phase velocity, generates both temporal transmission and temporal reflection. In contrast to the spatial mirror, the temporally reflected contribution enforces anti-causal temporal ordering, aligned with time reversal, with the wake tail preceding the refocusing front. Nevertheless, perfect reconstruction is still not achieved. First, the reflection operator intrinsically involves a time-derivative action, so the reflected waveform is not an exact time-reversed copy of the original impulse response even at the level of a single Fourier component. Second, the wake-tail contribution requires accounting for secondary radiation as the inward field reaches the origin. In our formulation this effect enters through the additional convolution term in \eqref{solution-time-mirror}, whose source function \(s(t)\) is obtained from an ill-conditioned Abel-type inverse problem. The numerical stabilization (finite-radius matching and Tikhonov regularization) is therefore not merely a technical choice but reflects the underlying sensitivity of attempting to reconstruct tail-dominated dynamics near the refocusing time.

Taken together, these results clarify that imperfect refocusing in two dimensions can persist even in idealized, lossless settings with perfect boundaries and complete control over the medium parameters. The limitation is therefore not primarily due to aperture truncation, attenuation, or measurement noise, but instead arises from the intrinsic temporal memory of even-dimensional propagators. From a practical standpoint, this implies that two-dimensional and quasi-two-dimensional platforms (e.g., surface waves, thin plates, planar waveguides, and effectively 2D electromagnetic or acoustic environments) may exhibit a systematic broadening of refocused pulses that cannot be eliminated simply by improving hardware or increasing measurement coverage.

Strategies aimed at mitigating wake-tail effects have been explored through dispersion engineering and the use of metamaterials designed to reshape wave propagation \cite{Bender}. However, introducing controlled dispersion generally involves a trade-off: while it may help compress or redistribute energy in time, it also modifies the propagation dynamics that refocusing techniques rely upon. In applications such as imaging or localization, where faithful reconstruction of the propagation channel is essential, additional dispersion can interfere with accurate refocusing and potentially degrade reconstruction fidelity. Consequently, suppression of wake-tail effects through material engineering must be balanced against the need to preserve the propagation characteristics.

In conclusion, the wake-tail structure inherent to two-dimensional wave propagation imposes a fundamental constraint on wave concentration. We find that, even under idealized conditions employing either spatial or temporal reflection, perfect refocusing remains unattainable. These results demonstrate that dimensionality alone introduces limits and should therefore be explicitly considered when interpreting relevant experiments and designing wave-control strategies in effectively two-dimensional systems.

\appendix

\section{\label{sec:Append}Proof of recursive relation}

For any sufficiently smooth radial function $f(r)$, the radial Laplacian in $d$ dimensions can be written as

\begin{equation}
{\Delta _d}f = \frac{1}{{{r^{d - 1}}}}\frac{\partial }{{\partial r}}\left( {{r^{d - 1}}\frac{{\partial f}}{{\partial r}}} \right),
\label{eq:app1}
\end{equation}

while the "dimension-raising" operator is defined as: $\mathcal{D} =   - \frac{1}{{2\pi r}}{\partial _r}$. The commutative property: ${\Delta _{d + 2}}\mathcal{D} = \mathcal{D}{\Delta _d}$ holds for radial functions, since:

\begin{equation}
{\Delta _{d + 2}}\mathcal{D}f(r) = \frac{1}{{{r^{d + 1}}}}\frac{\partial }{{\partial r}}\left( {{r^{d + 1}}\frac{\partial }{{\partial r}}\left[ { - \frac{1}{{2\pi r}}\frac{{\partial f}}{{\partial r}}} \right]} \right) =  - \frac{1}{{2\pi r}}\frac{\partial }{{\partial r}}\left( {\frac{1}{{{r^{d - 1}}}}\frac{\partial }{{\partial r}}\left( {{r^{d - 1}}\frac{{\partial f}}{{\partial r}}} \right)} \right) = \mathcal{D}{\Delta _d}f(r).
\label{eq:app2}
\end{equation}

(\ref{eq:app2}) is the appropriate operator to prove the recursive relation. ${g_d}(r,t)$ is the retarded Green function of the wave equation in $d$ spatial dimensions and the time derivative commutes with $\mathcal{D}$, since $\mathcal{D}$ acts only on $r$. Therefore

\begin{equation}
\left( {\frac{1}{{{c^2}}}\partial _t^2 - {\Delta _{d + 2}}} \right)\mathcal{D}{g_d} = \mathcal{D}\left[ {\left( {\frac{1}{{{c^2}}}\partial _t^2 - {\Delta _d}} \right){g_d}} \right] = \mathcal{D}\left[ {\delta (t){\delta ^{(d)}}({\bf{r}})} \right] = \delta (t){\delta ^{(d + 2)}}({\bf{r}}),
\label{app4}
\end{equation}

where the last step uses the standard distribution identity: $\mathcal{D}\left[ {{\delta ^{(d)}}({\bf{r}})} \right] = {\delta ^{(d + 2)}}({\bf{r}})$.

Since $\mathcal{D}$ preserves causality, $\mathcal{D}g_d$ is retarded. Therefore:

\begin{equation}
{g_{d + 2}}(r,t) = \mathcal{D}{g_d}(r,t) =  - \frac{1}{{2\pi r}}\frac{\partial }{{\partial r}}{g_d}(r,t).
\label{app5}
\end{equation}

\section{Temporal transmission and reflection at an instantaneous time mirror}

\subsection{Derivation of $T(\omega)$ and $R(\omega)$}

Consider a monochromatic component present just before the temporal modulation at $t=\tau_0$, $u^{-}(t)=e^{-i\omega (t-\tau_0)}$.
After the instantaneous event, the medium parameters return to their original values, so the field must be a superposition of forward- and backward-frequency components, $u^{+}(t)=T(\omega)e^{-i\omega (t-\tau_0)}+R(\omega)e^{+i\omega (t-\tau_0)}$.

Continuity of the field \eqref{boundary-relations1}, gives $T(\omega)+R(\omega)=1.$ Using $\partial_{t} u^-(\tau_0)=-i\omega$ and $\partial_{t}^{2}u^-(\tau_0)=-\omega^2$, while $\partial_{t}u^+(\tau_0)=-i\omega T(\omega)+i\omega R(\omega),$ the jump condition \eqref{boundary-relations2} yields $-i\omega T(\omega)+i\omega R(\omega)+i\omega=-\alpha\omega^2.$ Solving this system gives $T(\omega)=1-\frac{i\alpha\omega}{2}$ and $R(\omega)=\frac{i\alpha\omega}{2}.$

The coefficient $T(\omega)$ corresponds to temporal transmission, while $R(\omega)$ generates the time-reversed component.

\subsection{Time-domain operator form}

For a generic temporal waveform $f(t)$, multiplication of a Fourier component by $-i\omega$ corresponds to the time derivative $\partial_t$. Hence, the transmitted and reflected contributions may be written in the time domain as
\[
\mathcal{T}f(t)
=H(t-\tau_0)\left(f(t)+\frac{\alpha}{2}\partial_t f(t)\right),
\]
\[
\mathcal{R}f(t)
=H(t-\tau_0)\,\frac{\alpha}{2}\partial_t f(2\tau_0-t).
\]
The operator $\mathcal{T}$ describes the transmitted forward field, while $\mathcal{R}$ generates the time-reversed component emitted at $t=\tau_0$. Applying these operators to a monochromatic component
\[
f(t)=e^{-i\omega (t-\tau_0)}
\]
gives
\[
\mathcal{T}f(t)
=
H(t-\tau_0)
\left(1-\frac{i\alpha\omega}{2}\right)
e^{-i\omega (t-\tau_0)},
\]
and
\[
\mathcal{R}f(t)
=
H(t-\tau_0)
\left(\frac{i\alpha\omega}{2}\right)
e^{+i\omega (t-\tau_0)},
\]
which reproduces exactly the coefficients $T(\omega)$ and $R(\omega)$ above.

\section{Finite-radius matching for the calculation of $s(t)$ of last term of \eqref{solution-time-mirror-regular}}

\eqref{source-relation} leads to an ill-conditioned inverse problem, which becomes numerically unstable, particularly in the vicinity of $t=2\tau_0$. To bypass this difficulty numerically, we instead match the field on a small circle of radius $b \ll c{\tau _0}$ and impose $u_{ref}(b,t)=h(t-b/c)$, where the delay $b/c$ ensures causality. Therefore,

\begin{equation}
u_{ref}(b,t)
=
\int_0^t
s(\tau)\,
\frac{H\!\left(t-\tau-\frac{b}{c}\right)}
{2\pi \sqrt{(t-\tau)^2-\left(\frac{b}{c}\right)^2}}
\, d\tau,
\end{equation}

so that the trace at $r=b$ satisfies the causal Volterra equation of the first kind, $y(t) = \int_0^t {g(t - \tau )}{s(\tau )d\tau ,} $ where $g(t) = {{H(t - {b \mathord{\left/
 {\vphantom {b c}} \right.
 \kern-\nulldelimiterspace} c})} \mathord{\left/
 {\vphantom {{H(t - {b \mathord{\left/
 {\vphantom {b c}} \right.
 \kern-\nulldelimiterspace} c})} {\left( {2\pi \sqrt {{t^2} - {{\left( {{b \mathord{\left/
 {\vphantom {b c}} \right.
 \kern-\nulldelimiterspace} c}} \right)}^2}} } \right)}}} \right.
 \kern-\nulldelimiterspace} {\left( {2\pi \sqrt {{t^2} - {{\left( {{b \mathord{\left/
 {\vphantom {b c}} \right.
 \kern-\nulldelimiterspace} c}} \right)}^2}} } \right)}}$ and $y(t) = h({{t - b} \mathord{\left/
 {\vphantom {{t - b} c}} \right.
 \kern-\nulldelimiterspace} c})$. We discretize time uniformly, $t_{i}=i\Delta t$ for $i=0,...,N$ and approximate the convolution integral by a quadrature rule, yielding a lower-triangular linear system, ${y_i} \approx \sum\nolimits_{j = 0}^i {{A_{ij}}{s_j},} $, with ${A_{ij}} = \Delta tg({t_i} - {t_j})$ for $j \le i$ and $A_{ij}=0$ for $j > i$. The kernel $g(t)$ contains an integrable square-root singularity at $t=b/c$, making the associated operator compact and the inversion ill-conditioned. To stabilize the inversion, we compute $s$ using Tikhonov regularization \cite{Hansen}, $s = \arg \mathop {\min }\limits_s \left( {\left\| {As - y} \right\|_2^2 + {\lambda ^2}\left\| s \right\|_2^2} \right)$, where $\lambda$ is selected as a small fixed parameter that ensures numerical stability while maintaining good agreement between the reconstructed and prescribed waveforms.

\begin{acknowledgments}
 This work was supported by the “Stamatis G. Mantzavinos” Postdoctoral Scholarship of the Bodossaki Foundation.
\end{acknowledgments}

\bibliography{apssamp.bib}

\end{document}